\documentstyle[twocolumn,prl,aps]{revtex}

\begin{document}

\twocolumn[\hsize\textwidth\columnwidth\hsize\csname
@twocolumnfalse\endcsname

\title{``Incerto tempore, incertisque loci'':\\ Can we 
compute the exact time at which a quantum measurement 
happens?}
\author{\bf Carlo Rovelli}\address{Physics Department, 
University of Pittsburgh, Pa 15260, USA} \address{rovelli@pitt.edu} 
\date{\today} 
\maketitle
\begin{abstract}
Without addressing the measurement problem (i.e.\,{\em what\/} 
causes the wave function to ``collapse'', or to 
``branch'', or a history to become realized, or a property 
to actualize), I discuss the problem of the {\em 
timing\/} of the quantum measurement: assuming that in an 
appropriate sense a measurement happens, {\em when\/} 
precisely does it happen?  This question can be posed 
within most interpretations of quantum mechanics.  By 
introducing the operator $M$, which measures whether or not 
the quantum measurement has happened, I suggest that, 
contrary to what is often claimed, quantum mechanics does 
provide a precise answer to this question, although a 
somewhat surprising one.

\end{abstract}

\vskip2pc]

\vskip2cm

\noindent Quantum mechanics was not originally formulated as 
a general dynamical theory of the world, but as the theory 
of quantum microsystems interacting with classical 
macrosystems.  In the original formulation of the theory, 
the interaction of a quantum microsystem $S$ with a 
classical macrosystem $O$ is described in terms of ``quantum 
measurements''.  If the macrosystem $O$ interacts with the 
variable $q$ of the microsystem $S$, and $S$ is in a 
superposition of states with different values of $q$, then 
the macrosystem $O$ ``sees'' only one of the values of $q$, 
and the interaction modifies the state of $S$ by projecting 
it into a state with that value.

It was noticed early that this formulation raises certain 
difficulties, particularly if we want to view the theory as 
the general dynamical theory of anything, and not just 
microsystems in the laboratory.  These difficulties are 
usually referred to as the ``measurement problem'' and they 
have motivated ingenious attempts to modify either quantum 
mechanics or its interpretation.

In this letter, I do not discuss the measurement problem.  
Rather, I discuss a secondary problem, which as far as I 
understand, appears in most (but not all) formulations of 
quantum mechanics.  The problem I discuss is the 
determination of the precise time at which the quantum 
measurement happens.  Let me illustrate, without any 
ambition of precision or completeness, how this problem 
manifests itself in some popular interpretations of the 
theory.  In one interpretation, a system $S$ has a wave 
function, which collapses during a measurement.  Is the 
collapse of the wave function instantaneous?  If so, when 
precisely does it happen?  In another interpretation, the 
system is described in terms of its properties, or values of 
its dynamical variables.  These become manifest, and take 
definite values when observed, in a measurement.  Do 
properties become manifest suddenly?  When precisely do they 
become manifest, in a realistic laboratory experiment?  In 
another interpretation, the wave function never collapses 
but it branches.  When precisely does the branching happens?  
In another interpretation, the wave function never branches, 
but we, observers, navigate through it and sit in one or the 
other of its components.  When is it, precisely, that the 
selection of this or that component happens?  In another 
interpretation, there is no wave function, but only 
probabilities assigned to sequences of (quantum) events.  
When is it precisely that a quantum event ``happens''?  
Namely, when precisely can we replace the statement ``this 
may happen with probability $p$'' with the statement ``this 
has happened''?

The most common answer to these questions is that quantum 
mechanics does not determine the time at which the 
measurement happens.  As is well known, for instance, von 
Neumann, pointed out that we can freely move the boundary 
between the quantum system and the observer, and therefore 
also shift the measurement's presumed time, without 
affecting physical predictions \cite{vn}; and Heisenberg 
insisted that we can compute probability transitions between 
initial and final states, but we must abstain from asking 
``in between'' questions \cite{h}.  According to this view, 
the question ``when does the measurement happen'' does not 
seem to be a well posed question.

In this letter, I claim that, contrary to that view, quantum 
mechanics {\em does\/} give a precise answer to this 
question, although a peculiar answer.  More precisely, I 
claim that: (i) a precise (operational) sense can be given 
to the question of the timing of the measurement; (ii) we 
can then compute the time at which the measurement happens 
(in the sense specified) using standard quantum techniques; 
(iii) the (more or less realistic) interpretation of the 
physical meaning of this time is no more problematic than 
the interpretation of any other quantum result.  I will also 
venture in suggesting that this observation might partially 
illuminate some aspects of the major problem -- the measurement 
problem itself. 

The answer I present here is based on two main ideas.  
First, that the question ``When does the measurement 
happen?''  is quantum mechanical in nature, and not 
classical.  Therefore its answer must be probabilistic.  In 
general, in quantum mechanics the answer to a question such 
as ``Is the spin up?''  is not ``Yes'' or ``No'', but 
rather: ``Yes with probability, say, $1/2$'', which implies 
that the spin will come out ``up'' in half the repetitions 
of the experiment.  Similarly, I am roughly going to argue 
that ``half way through a measurement'' the measurement is 
not ``partially happened'', but just ``happened with 
probability 1/2'', or ``already realized in half of the 
repetitions of the experiment''.  The second idea is that 
the question ``When does the measurement happen?''  does not 
regard the measured quantum system $S$ alone, but rather the 
coupled system formed by the observed system $S$ and the 
observer system $O$.  Therefore the appropriate theoretical 
setting for answering this question is the quantum theory of 
the two coupled systems.  In particular, I will submit that 
there is no contradiction in using the quantum theory of $S$ 
for describing the observed behavior of $S$ and the quantum 
theory of the coupled $S-O$ system for answering the 
question of the precise timing of the measurement.

More in detail, this paper is based on a technical 
observation: there is an operator that has a natural 
interpretation as measuring whether or not the measurement 
has happened.  I will denote this operator as $M$ (for 
Measurement).  The operator $M$ is a projection operator, 
with the eigenvalue 1 meaning ``the measurement has 
happened'', and the eigenvalue 0 meaning ``the measurement 
has not happened''.  By applying standard quantum mechanical 
rules to {\em this\/} operator, at every time $t$ we can 
compute a precise (although probabilistic) answer to the 
question whether or not the measurement has happened.

The description of the quantum measurement that I employ 
here is based on the Peres-Wootters analysis of the quantum 
measurements of finite duration \cite{pw}.  Several of the 
ideas on which this paper is based were first introduced 
therein.  In particular, the key idea that measurement can 
be defined operationally, in terms of the correlation that 
could be detected by a {\em second\/} external apparatus, is 
in \cite{pw}.  Peres and Wootters do not introduce the 
operator $M$ considered here, but discuss the related idea 
of computing the probability distribution of the measured 
variable, conditionalized by the observed position of the 
apparatus' pointer.  The aim of Peres and Wootters was to 
provide a physical analysis of realistic measurements and
their intrinsic limitations.  Here, on the other hand, I am 
interested in the general notion of measurement and on what 
we can say on {\em when\/} we replace the statement ``this 
may happens with probability $p$'' with the statement ``this 
has happened''.

The result described here has emerged within the 
relational interpretation of quantum mechanics 
\cite{rovelli}.  See also the strictly related Kochen's 
interpretation \cite{kochen}, and reference \cite{relation}, 
where these two interpretations are compared.  This 
result, however, does not require such interpretations, and I 
present it here in a form which is independent from the 
interpretation of quantum mechanics one holds. 

\vskip.5cm

Just to fix a language, I will refer here to a traditional 
(and a bit out of fashion) ``wave function collapse'' 
terminology, leaving to the reader the burden of translating 
what I say in his or her preferred language.  Consider a 
physical system $S$ (an electron).  Assume that $S$ 
interacts with another physical system $O$ (an apparatus 
measuring the spin of the electron), and that the 
interaction between $S$ and $O$ qualifies as a quantum 
measurement of the variable $q$ of the system $S$.  Choosing 
the simplest setting, and following a terminology which is 
now standard, I assume that $q$ has only two discrete 
eigenvalues, $a$ and $b$, and denote by $|a\rangle$ and 
$|b\rangle$ the two corresponding eigenstates.  The 
interaction between $S$ and $O$ is governed by an 
interaction hamiltonian $H_{I}$.  A necessary condition for 
the interaction to be a measurement is that we can prepare 
$O$ in an initial state, which we denote $|init\rangle$, 
such that in a finite time $T$ the interaction will evolve 
(perhaps up to some approximation) 
$|a\rangle\otimes|init\rangle$ into 
$|a\rangle\otimes|Oa\rangle$ and 
$|b\rangle\otimes|init\rangle$ into 
$|b\rangle\otimes|Ob\rangle$, where $|Oa\rangle$ and 
$|Ob\rangle$ are states of $O$ that we can identify as ``the 
pointer of the apparatus indicates that $q$ has value $a$'', 
or $b$, respectively.  Up to now, I have only given 
definitions.  As it is well known, the linearity of quantum 
mechanics implies that if $S$ is initially in a quantum 
superposition $c_{a}|a\rangle+c_{b}|b\rangle$, the combined 
system will evolve from
\begin{equation}
	\Psi(0)=\big(c_{a}|a\rangle 
+c_{b}|b\rangle\big)\otimes| init\rangle
	\label{prima}
\end{equation}
into 
\begin{equation}
	\Psi(T) = 
c_{a}|a\rangle\otimes|Oa\rangle + 
c_{b}|b\rangle\otimes|Ob\rangle. 
	\label{dopo}
\end{equation}
Examples of models of interactions of the kind described are 
well known (see for instance \cite{measurements}), and are 
easy to construct in the case of simple experiments such as 
a Stern-Gerlach measurement.\footnote{The difficulty of 
describing the dynamics of quantum measurement in terms of 
Schr\"odinger evolution is not in showing how the system may 
evolve into a correlated state.  It is in understanding how 
it may go from the correlated state to a mixture, or to an 
eigenstate of the measured quantity.}

For some reason, at some point we have to (or we can) 
replace the pure state $\Psi(T)$ with a mixed state.  
Equivalently, we replace $\Psi(T)$ with either
\begin{equation}
\Psi_{a}=|a\rangle\otimes|Oa\rangle,	
	\label{a}
\end{equation}
 or 
\begin{equation}
\Psi_{b}=|b\rangle\otimes|Ob\rangle,	
	\label{b}
\end{equation}
where, of course, the probability of having one or the other 
is $|c_{a}|^{2}$ and $|c_{b}|^{2}$ respectively.  Since here 
I am not discussing the measurement problem, I will not 
address the problem of the meaning of this step (whether it 
is a physical event, a change in our knowledge, a mental 
event, a perspectival event, or other).  Nor I will discuss 
whether, or how, or why, or under which circumstances, this wave 
function collapse happens.

Rather, I simply assume that in some appropriate sense the 
wave function changes from (\ref{dopo}) to either 
(\ref{a}) or (\ref{b}) and the quantity $q$ acquires a 
definite value; and I focus on the question of what we can 
say about the precise time $t$ at which this happens.

At the time $T$, the system has reached the state 
(\ref{dopo}), and a definite correlation between the pointer 
variable, with eigenstates $|Oa\rangle$ and 
$|Ob\rangle$, and the system variable, with eigenstates 
$|a\rangle$ and $|b\rangle$, is established.  Before that 
time, in general, this correlation is absent, or incomplete.  
If the wave function has collapsed, and the state is either 
(\ref{a}) or (\ref{b}), the correlation is present as well.  
In other words, the happening of the measurement is tied to 
the fact that the complete correlation is established.

This well known observation has prompted many to suggest that the 
measurement (of the quantity $q$, by the pointer variable) 
happens in the moment in which the complete correlation is 
established.  In modal interpretations, for instance, the 
quantity $q$ ``has value'' only if the state is in the (Schmidt, 
or biorthogonal) form (\ref{dopo}).  This would lead us to say 
that the measurement of $q$ happens instantaneously in the moment 
the complete correlation is established.\footnote{This solution 
does not convince me for various reasons.  First, nature is never 
clean, and in a realistic case the exact form (\ref{dopo}) is 
never attained.  One might reply that if the state is ``close'' 
to the form (\ref{dopo}), then it will still have a Schmidt 
decomposition into a basis ``close'' to the eigenbasis of $q$; 
but unfortunately this is not true: Schmidt bases jump badly 
anytime the state passes close to a degenerate state, and a state 
``very close to (\ref{dopo})'' may defines a Schmidt basis in 
which a quantity completely unrelated to $q$ is diagonal.  
Second, quantum mechanics gives us probabilistic statements about 
nature.  Why should it give us a sharp answer for the measurement 
time?} I consider here a different approach to the problem.

Notice that there is a precise sense in which we can {\it 
measure\/} whether, at some time $t$ intermediate between 
$t=0$ and $t=T$, the measurement has happened or not.  In 
fact, let us suppose that at $t$ we bring a further external 
apparatus $O'$ in and we measure the value of $q$ as well as 
the position of the pointer (the two commute, of course).  
Then, either ({\it Case 1\/}) we find the pointer in the 
correct position corresponding to the value of $q$ that we 
have found - namely we obtain \{$a$ and $Qa$\}, or \{$b$ and 
$Qb$\}; or ({\it Case 2\/}) the pointer is not on the correct 
position.\footnote{Typically, the pointer 
will fail to be in the correct position not because it is in the 
wrong position, but because it is still in $|init\rangle$.}
The question ``Has the pointer already measured 
$q$ at time $t$?''  has therefore a natural 
interpretation: if we find {\it Case 1}, we can say that the 
pointer has already measured $q$; if we find {\it Case 2}, 
it hasn't.  Thus, the question of the timing at which a 
measurement happens can be given a simple operational 
interpretation, independent from any metaphysics, or any 
interpretation of quantum mechanics, one might hold.  The 
operational procedure described assigns a precise meaning to 
the question.  Can we then answer the question?  Certainly 
yes, because quantum mechanics provides us the tool for 
predicting the outcome of the measurement described.  More 
precisely, it provides us the tool for computing the 
probabilities of finding {\it Case 1} or {\it Case 2}, for 
any given state $\Psi(t)$ of the combined $S-O$ system.  
These probabilities are obtained as follows.

Define the operator $M$ on the state space of the $S-O$ 
system, as the projection operator on the subspace spanned by 
the two states $\Psi_{a}$ and $\Psi_{b}$: 
\begin{equation}
	M \equiv 
	|\Psi_{a}\rangle\langle\Psi_{a}|+
	|\Psi_{b}\rangle\langle\Psi_{b}|. 
\end{equation}
Namely 
\begin{eqnarray}
	M \big(|a\rangle\otimes|Oa\rangle\big) 
	& = & |a\rangle\otimes|Oa\rangle 
\\
	M \big(|b\rangle\otimes|Ob\rangle\big) 
	& = & |b\rangle\otimes|Ob\rangle
\\
	M \big(|a\rangle\otimes|Ob\rangle\big) 
	& = & 0
\\
	M \big(|b\rangle\otimes|Oa\rangle\big) 
	& = & 0
\end{eqnarray}
and 
\begin{equation}
	M \big(|\psi\rangle\otimes|\phi\rangle\big) =  0
\end{equation}
if $\langle\phi|Oa\rangle=0$ and $\langle\phi|Ob\rangle=0$.  

$M$ is a self-adjoint operator on the Hilbert space of the 
coupled system.  Therefore it may admit an interpretation 
as an observable property of the coupled system.  In all the 
eigenstates of $M$ with eigenvalue 1 the pointer variable 
correctly indicates the value of $q$.  In all the 
eigenstates of $M$ with eigenvalue 0, it does not.  
Therefore, $M$ has the following interpretation: $M=1$ means 
that the pointer (correctly) measures $q$.  $M=0$ means that 
it does not.  Now, when the pointer of the apparatus 
correctly measures the value of the observed quantity, we 
say that the measurement has happened.  Therefore we can say 
that $M=1$ has the physical interpretation ``The measurement 
has happened'', and $M=0$ has the physical interpretation 
``the measurement has not happened''.

Since $M$ is a genuine self-adjoint operator on the Hilbert 
space of the coupled system, and since it has an unambiguous 
physical interpretation, we can apply the standard 
interpretation rules of quantum mechanics to it.  In 
particular, for all states that are not eigenstates of $M$, 
quantum mechanics tells us that we must not interpret 
these states as having ``intermediate'' values; but, 
rather, they have one or the other values with a 
certain probability.  That is, if a state $\Psi$ of the 
coupled system is not a perfectly correlated state of the 
form (\ref{dopo}), (\ref{a}), or (\ref{b}), then in 
general we must conclude that $\Psi$ is a superposition of a 
state in which the measurement has happened and a state in 
which the measurement has not happened.

In particular, if we follow the Schr\"odinger evolution 
$\Psi(t)$ of the state of the coupled system from $t=0$ to 
$t=T$, then at every intermediate $t$ we can compute the 
probability $P(t)$ that the measurement has happened
\begin{equation}
	P(t)=\langle \Psi(t)|M|\Psi(t) \rangle
	\label{p}
\end{equation}
For a good measurement, $P(t)$ will be a smooth function 
that goes monotonically from 0 to 1 in the time interval 0 
to $T$.  Therefore, half way through the measurement, we 
cannot say that the measurement has ``half happened'': we 
must say that the measurement ``has happened with 
probability $P(t)$''.

The last statement has a precise operational meaning: we may 
repeat the physical process in which $O$ measures the 
quantity $q$ of $S$ a large number of trials.  In each 
trial, at a time $t<T$ after the beginning of the process, 
we can check whether or not the measurement has already 
happened by having the second ``external'' apparatus $O'$ 
measuring $q$ and the position of the pointer.  If the two 
match, then $O$ has detected the ``correct'' value of $q$:  
``a measurement has happened''.  Standard quantum mechanical 
arguments show that $P(T)$, defined in (\ref{p}), gives 
precisely the fraction of the trials in which we will detect 
that the measurement has happened. 

The correctness of this interpretation of $M$ is 
particularly evident in the following case.  Imagine that 
the two pointer states $|Oa\rangle$ and $|Ob\rangle$ 
represent two LED's that light up when the $O$ 
apparatus has measured $a$, or, respectively, $b$.  Also, 
assume that in the measurement the ``wrong'' states 
$|a\rangle\otimes |Ob\rangle$ and $|b\rangle\otimes 
|Oa\rangle$ happens with null or negligible probability (a 
realistic assumption for a good measurement).  If we ask an 
experimentalist when is the measurement completed, he 
will probably answer: ``when one of the LED's lights up''.  
But the distribution probability of the time at which the 
LED lights up can be theoretically computed, and is 
given precisely by $P(t)$, because $M$ is the projector on 
the subspace on which one of the LED's is on.

If $P(t)$ is the probability that the measurement has 
happened at time $t$, then the probability (density) $p(t)$ 
that the measurement happens between time $t$ and time $t+dt$ 
is 
\begin{eqnarray}
	p(t) &=& \frac{dP(t)}{dt} \nonumber \\
	&=&\frac{d}{dt}\ \langle \Psi(t)|M|\Psi(t) \rangle \nonumber \\
	&=&	\langle \Psi(t)| [M,H] |\Psi(t) \rangle,
\end{eqnarray}
where $H$ is the total hamiltonian. I introduce the operator 
\begin{equation}
	m = [M,H]. 
\label{m}
\end{equation}
Its expectation value at time $t$ gives the probability density for 
the measurement to happen at time $t$
\begin{equation}
		p(t) =	\langle \Psi(t)| m |\Psi(t) \rangle. 
	\label{q}
\end{equation}
For a good measurement in which $P(t)$ grows smoothly and 
monotonically from zero to one, $p(t)$ will a ``bell 
shaped'' curve, defining the time at which the measurement 
happens, and its quantum dispersion. 

The operators $M$ and $m$ can be generalized to an arbitrary number 
of eigenvalues of the quantity measured.  If the apparatus 
is able to distinguish the (eigen)values $a_{1}\ldots a_{n}$ 
of the variable $q$, by means of the pointer positions 
$Oa_{1}\ldots Oa_{n}$, then $M$ is given by
\begin{equation}
	M = \sum_{i=1,n} \big(|a_{i}\rangle\otimes|Oa_{i}\rangle\big)
	\big(\langle a_{i}|\otimes\langle Oa_{i}|\big),
	\label{genfor}
\end{equation}
and this definition can easily be generalized to continuous 
spectra. 

Thus, we can conclude that quantum mechanics provides a 
precise meaning and a precise answer to the question of when 
a measurement happens.  The answer is, like any other answer 
to physical questions in quantum theory, a probabilistic 
one.\footnote{Quantum events, the individual events at the 
basis of modern physics, happen randomly -- precisely as the 
Democritean $\kappa\lambda\iota\nu\grave{\alpha}\mu\epsilon\nu$, 
the individual events at the 
basis of ancient physics, which, in Lucretius marvelous 
verses, happen ``incerto tempore \ldots  incertisque loco'': at a 
random time and in a random place.\cite{lu}} Using the 
operator $m$ defined in (\ref{m},\ref{genfor}), the probability 
$p(t)$ that the measurement happens at time $t$ can be computed at every 
intermediate time from (\ref{q}). This quantity can be 
explicitly computed from the standard quantum physics of 
the system. 

\vskip1.3cm 

I close with some observations and general comments.  
The first observation refers to the relation between the issue 
discussed here and the general measurement problem.  The 
interpretation of the operator $M$ and of the predictions it 
yields are infested by the same subtle problems that infest any 
other quantum operator.  For instance, one may ask whether $M$ 
measures ``if the measurement has indeed really happened'', or 
whether ``we shall find that the measurement has happened if we 
check''.  The distinction refers to the very possibility of 
assigning values to quantities irrespectively from observations, 
and thus touches the core of the measurement problem.  As I said, 
I do not discuss this major conundrum here.  The interpretation 
of the time at which a measurement happens has no less and no 
more difficulties than the interpretation any other quantity of a 
quantum system. In particular, I warn the reader of not 
charging my claim with more meaning than what it deserves.  I am 
not claiming that there is an ``element of reality'' in the fact 
that a measurement has happened, or that, in general 
``measurement having happened or not" is an {\it objective 
property} of the coupled system.  I am only claiming: i) that the 
the time $t$ at which a measurement happens is a well defined 
concept in the theory (so precise that it can be defined 
operationally); and ii) that this time, or, more precisely its 
probability distribution, can be explicitly computed if the 
unitary dynamics of the system-apparatus dynamics is known. 
Thus, this time is as well defined a quantity as, say, the 
position of a particle: if we know the dynamics, we can, at any 
moment, compute its probability distribution. 

The notion of measurement that I have referred to in this paper 
does not require that phase coherence is definitely lost.  Also, 
I have made no reference to physical decoherence.  If quantum 
mechanics is {\em exactly\/} correct in nature (as I assume 
here), then phase coherence is never definitely lost.  (In 
principle one can always make an interference experiment between 
the spin up and the spin down components of the state, even after 
a macroscopic Stern Gerlach apparatus has registered the outcome 
on printed paper.)  In practice, phase coherence is lost via 
physical decoherence, say into the environment, as beautifully 
realized by Zurek \cite{decoherence}, making the pure and the 
mixed states effectively indistinguishable.  Thus, the question 
of when a quantum correlation (the pure state) can be {\em 
effectively\/} described as a mixed state is already solved.  
This, however, is not the question I have addressed here.  The 
question I have addressed refers to the fact that one or the 
other of the physical values of $q$ become, at some point, part 
of our reality: a mystery that, as Zurek emphasizes in the 
conclusion of \cite{decoherence}, remains also after 
understanding physical decoherence.  Without addressing the issue 
of how this actualization of the values of $q$ may come about, I 
have discussed here ``when'' we may say it happens.

An issue often discussed in the context of the measurement 
problem is the one of imperfect measurements. Imagine that an 
interaction between a system $S$ and an ``apparatus'' $O$ is such 
that after the interaction the coupled system ends up in a state 
which is close, but not exactly equal, to the state (\ref{dopo}).  
Has the measurement happened, even if perfect correlation has not 
been established?  The approach presented here offers a solution 
to the conundrum.  The solution is that the measurement has 
happened with high probability.  In practical terms, this means 
that it has happened in almost all, but not all, the trials.

A related observation is that the existence of the $M$ 
operator might perhaps cause an embarrassment for the modal 
interpretation.  In fact, quantum mechanics contains an 
operator corresponding to the question ``When does $q$ take 
a definite value?'', a question addressed in detail within 
modal views.  But the answer that quantum mechanics provides 
seems inconsistent with the modal view that $q$ has a 
definite value only when the requisite biorthogonal 
correlation is established.\footnote{I thank John Earman for 
this observation.}

A possible objection to the present work is that it is 
relevant within certain interpretations of quantum 
mechanics, but it is irrelevant within the interpretations 
in which no mention of quantum measurement is made, such as 
the histories interpretations \cite{histories}.  This is a 
delicate issue, which requires a case by case discussion 
that would not fit into this brief note.  I only sketch here 
some general considerations on this regard.  Following 
\cite{pw}, I have given an operational definition of the 
measurement's timing.  Under such a definition, 
measurement's timing makes sense within any interpretation.  
But why would we want to discuss measurement within a 
measurement-free interpretation?  As suggested in the 
beginning of the paper, I think that within most, if not 
all, interpretations, there is always a point in which we 
have to jump from the statement that something {\em may\/} 
happen with probability $p$, to the statement that 
something, in some appropriate sense, {\em has\/} happened.  
Otherwise, what is it that distinguishes the (probabilistic) 
{\em predictions\/} we {\em make\/} from the 
(non-probabilistic) {\em data\/} we {\em have\/} about the 
world, on which those predictions are based?  As cleanly 
made clear in particular by the histories interpretation, 
quantum mechanical predictions do not depend just on the 
data: they also depend on the question asked.  One can 
refuse to discuss the step in which predictions become data, 
as prescribed in the histories interpretation, and reset the 
theoretical machinery at every newly acquired data.  This is 
fine.  But nothing then prevents somebody else to ask the 
further question of whether one can predict if and when the 
theoretical machinery can be reset.  The 
histories-interpretation view, I think, is that this 
question has no answer within quantum mechanics.  This is 
precisely Heisenberg's claim, mentioned above.  Here, 
contrary to that claim, I have suggested that one can still 
keep the interpretation as it is, but {\em add\/} to it, 
coherently, a meaningful prediction on when the resetting 
of the theoretical machinery can be made.\footnote{ I add 
a more speculative remark.  I suspect that in spite of 
brilliant attempts to purge quantum mechanics of any 
``special moment'' in time in which measurements happen, 
such ``special moments'' remain in most, if not all, 
interpretations of quantum mechanics, sometimes in a hidden 
fashion.  A coherent way of avoiding this problem is to 
relegate it to a still to be discovered theory of 
consciousness, as David Mermin has recently, coherently, 
suggested in \cite{mermin} (on a related vein, see 
\cite{saunders}).  Short of this noble and courageous 
declaration of failure, we may purge quantum mechanics from 
the expression ``measurement'', pleasing Bell \cite{bell}, 
but the mystery posed by these ``special moments'', whether 
we call them measurements, quantum events, or otherwise, or 
we refuse to name them, remains.  Talking about their 
timing, as I have attempted here, is an indirect way of 
addressing the issue.} \newpage

The answer to the timing problem I have discussed is 
peculiar in one respect: it involves the simultaneous use of 
the formal apparatus of quantum mechanics at two levels.  I 
have combined the use of the quantum theory of the $S-O$ 
system, in which the operator $M$ is defined, with the 
quantum theory of the system $S$ alone, in which we may talk 
of the collapse of the wave function due to the interaction 
of $S$ with $O$.  There is something slightly conceptually 
anomalous in doing that.  I think that this anomaly is not a 
defect of the idea proposed here.  Rather, it represents its 
essential aspect.  This procedure might shed some light on 
the relational aspects of quantum theory.  There is a subtle 
point here: the fact that a ``time of measurement'' can be 
computed does not imply that the wave function collapse is 
an objective observer-independent event.  On the contrary, 
to my understanding, the core of the measurement problem is 
the fact that if we take quantum mechanics as a general 
theory of mechanics, we must combine two apparently badly 
contradictory facts.  On the one hand, the value of the 
variable $q$ is definitely directly observed by $O$ (by us).  
On the other hand, phase coherence is never really lost, and 
thus the ``other branch'' is still somehow there, even if 
not {\em for\/} $O$ (for us).  The precise time at which the 
measurement happens can be observed, and can be 
operationally defined in the manner described above, but not 
by $O$.  Rather, by the third system ($O'$) making 
measurements on the $S-O$ system.  Simultaneous use of these 
different levels does not lead to any contradiction 
\cite{rovelli,kochen}.  And there is no regress at infinity, 
unless we inquire about absolute reality, an inquiry which, 
I suspect, is illegitimate.  These ideas are discussed, in 
various versions, in \cite{kochen} and in 
\cite{rovelli,relation}.  The existence of the $M$ operator 
and the solution of the timing problem suggested here find a 
natural framework within these views, but do not require 
them.  \vskip.5cm

I thank Rob Clifton for bringing reference \cite{kochen} to 
my attention and for a careful reading of this manuscript; 
Euan Squires, Jim Hartle and Bob Griffiths for a long and 
extremely valuable e-mail correspondence on these matters; 
John Earman, Chris Isham, John Baez, Bill Curry, Ted Newman, 
Lee Smolin and Louis Crane for comments and conversations.  
I also thank Asher Peres for bringing reference \cite{pw} to 
my attention, after I had posted the first version of this 
paper in the Los Alamos Archives.  I apologize with him and 
with Bill Wootters for having overlooked their important 
work at first.  Support for this work came from NSF grant 
PHY-95-15506.

\end{document}